# Optimal Band Allocation for Cognitive Cellular Networks

**Chengling Jiang, Tingting Liu**

*Abstract*—FCC new regulation for cognitive use of the TV white space spectrum provides a new means for improving traditional cellular network performance. But it also introduces a number of technical challenges. This letter studies one of the challenges, that is, given the significant differences in the propagation property and the transmit power limitations between the cellular band and the TV white space, how to jointly utilize both bands such that the benefit from the TV white space for improving cellular network performance is maximized. Both analytical and simulation results are provided.

## I. INTRODUCTION

IN recent years, cellular networks have transformed from providing mobile communication with limited data to supporting universal mobile broadband services. This has led to a large capacity demand that cannot be accommodated with the existing cellular spectrum resources. On the other hand, although radio spectrum is a scarce resource, it has been observed that licensed radio spectrum is not fully utilized everywhere at all times. Cognitive radios have been proposed as a solution to the spectrum crisis [1]. FCC has recently permits cognitive use of the TV spectrum (white space) [2]. One of the possible applications of the TV white space is to offload part of the cellular network throughput load to the white space. The 470-700 MHz white space provides superior propagation and building penetration compared to the band that the 3G and 4G cellular networks use (2-2.5 GHz). However, access to the TV white space for use in cellular communications also comes with some technical challenges. In this letter, we focus on studying the optimal joint use of cellular band and the TV white space for both overall system and individual user performance improvement. In Section II, we analyze the optimal band allocation for the downlink. That is, we derive a method of allocating cellular and TV band resources to different users such that both individual and overall system performance are maximized. In Section III, we study the optimal band allocation on the uplink. Simulation results are provided in Section IV, and Section V concludes this letter.

## II. OPTIMAL DOWNLINK BAND ALLOCATION

Assume that there are $N$ users, $\mathbf{U} \triangleq \{u, 1 \leq u \leq N\}$, served by a sector in a cellular network with transmit bandwidth $W_c$ and total power $P_c$ (e.g., 42 dBm plus 17 dBi antenna gain) in cellular band (e.g., 2 GHz). The achievable data rate for user $u \in \mathbf{U}$ in a traditional cellular network (no help from the TV white space spectrum) is

$$C(u) = \frac{W_c}{|\mathbf{U}|} \log\left(1 + \frac{P_c \eta_c(u)}{W_c N_0}\right) \quad (1)$$

where $W_c/|\mathbf{U}| = W_c/N$ is the cellular bandwidth allocated to each user, $\eta_c(u)$ is the path power gain in cellular band for user $u$ and $N_0$ is the noise spectral density.

Now we assume that some of the users in $\mathbf{U}$, i.e., $\mathbf{U}_T \subset \mathbf{U}$ are allocated to a vacant TV band with bandwidth $W_T$ and the rest of the users, $\mathbf{U}_c \triangleq \mathbf{U} \setminus \mathbf{U}_T$, stay in the cellular band. The data rate for a user in the cellular band, $u \in \mathbf{U}_c$, thus becomes

$$C_c(u) = \frac{W_c}{|\mathbf{U}_c|} \log\left(1 + \frac{\frac{P_c}{|\mathbf{U}_c|}\eta_c(u)}{\frac{W_c}{|\mathbf{U}_c|}N_0}\right) = \frac{W_c}{|\mathbf{U}_c|}\log\left(1 + \frac{P_c\eta_c(u)}{W_c N_0}\right) \quad (2)$$

$$= \frac{|\mathbf{U}|}{|\mathbf{U}_c|}C(u)$$

It is evident that whoever stays in the cellular band enjoys a $|\mathbf{U}|/|\mathbf{U}_c|$ times increase in data rate due to the increased bandwidth. As for the users who are moved over to the TV band, the new data rate is

$$C_T(u) = \frac{W_T}{|\mathbf{U}_T|}\log\left(1 + \frac{\frac{P_T}{|\mathbf{U}_T|}\eta_T(u)}{\frac{W_T}{|\mathbf{U}_T|}N_0}\right) = \frac{W_T}{|\mathbf{U}_T|}\log\left(1 + \frac{P_T\eta_T(u)}{W_T N_0}\right) \quad (3)$$

where $\eta_T(u)$ is the TV band path (power) gain of user $u \in \mathbf{U}_T$, $P_T$ is the maximum allowable transmit power on the TV frequency. By FCC, fixed devices are permitted to transmit up to 36 dBm. That is, $P_T = 36$ dBm, corresponding to a 23 dB loss in transmit power compared to the cellular band. This huge loss in transmit power in the TV band cannot be fully compensated by the advantage in path loss. Consequently, assuming the same bandwidth for both cellular and TV bands, i.e., $W_T = W_c = W$ (e.g., 5 MHz), a user may gain or lose data rate from the use of TV white space. There is an increase in data rate due to the increase of bandwidth by a factor of $|\mathbf{U}_T|$, i.e., the number of the users allocated to the TV band. However, depending on the user's geometry (i.e., position in the cell), this gain maybe offset by the potential drop in spectral efficiency as a result of the significant decrease in transmit power even with an increase in path gain. That is, not all users may benefit from the TV band.

We therefore look for a band allocation strategy that best utilizes the TV white space for improving both the individual user and the system performance. In particular, we seek $\mathbf{U}_T$ that maximizes a given objective function, typically, the proportional fair metric [3], i.e.,

$$\tilde{\mathbf{U}}_T = \arg\max_{\mathbf{U}_T} \left\{ \sum_{u \in \mathbf{U}_T} \log C_T(u) + \sum_{u \in \mathbf{U}_c} \log C_c(u) \right\} \quad (4)$$

$$= \arg\max_{\mathbf{U}_T} \left\{ \prod_{u \in \mathbf{U}_T} C_T(u) \prod_{u \in \mathbf{U}_c} C_c(u) \right\}.$$

The resource allocation scheme based on the proportional fair enables optimal tradeoff between each individual user's performance and the system performance as a whole. Unfortunately, an exhaustive search for the optimal $\mathbf{U}_T$ among all possible combinations of users in $\mathbf{U}$ can be computationally prohibitive. Instead, consider two arbitrary users in cellular band, $u_1, u_2 \in \mathbf{U}_c$, $\forall \mathbf{U}_c \subset \mathbf{U}$, at geometry $d_1$ and $d_2$, respectively, with $d_1 < d_2$, i.e., $u_1$ is closer to base station (higher geometry) than $u_2$. Assume we are to move one user from the cellular band to the TV band. Scheme 1 moves $u_1$ with higher geometry to the TV band ($u_2$ with lower geometry is maintained in the cellular band). Scheme 2 on the contrary moves $u_2$ to the TV band ($u_1$ remains in the cellular band). The corresponding proportional fair metrics for the two schemes are

$$\prod_{u \in \mathbf{U}_T + u_1} C_T(u) \prod_{u \in \mathbf{U}_c \setminus u_1} C_c(u) \qquad (5)$$

for the scheme 1, and

$$\prod_{u \in \mathbf{U}_T + u_2} C_T(u) \prod_{u \in \mathbf{U}_c \setminus u_2} C_c(u) \qquad (6)$$

for the scheme 2. The only terms that differentiate the above two proportional fair metrics are:

$$\log\left(1 + \frac{P_T \eta_T(u_1)}{WN_0}\right) \log\left(1 + \frac{P_c \eta_c(u_2)}{WN_0}\right) \qquad (7)$$

in (5), and

$$\log\left(1 + \frac{P_T \eta_T(u_2)}{WN_0}\right) \log\left(1 + \frac{P_c \eta_c(u_1)}{WN_0}\right) \qquad (8)$$

in (6), where $P_c = 59$ dBm and $P_T = 36$ dBm per FCC rules.

It can be verified that (7) is greater than (8). Consequently, (5) is greater than (6). That is, it is better (larger proportional fair metric) to move the higher geometry user to the TV band. We therefore come to a strong conclusion that, if $\tilde{\mathbf{U}}_T$ maximizes (4), it must contain the highest geometry users and $\tilde{\mathbf{U}}_c = \mathbf{U} \setminus \tilde{\mathbf{U}}_T$ the lowest geometry users.

This conclusion makes sense in that, comparing (3) with (1), if a user is ever to gain a data rate increase from the TV band, it is more likely the higher geometry user who have larger path gains than the low geometry users to compensate for the loss in transmit power.

The significance of this conclusion is that it simplifies the optimization process in (4) to the much more manageable optimization problem of determining $\left|\tilde{\mathbf{U}}_T\right|$, i.e., the number of top highest geometry users to be moved to the TV band that maximizes the proportional fair metric. This problem can be easily implemented by "hypothesis-test-moving" the highest geometry users to the TV band one by one from the top of the pre-sorted user list until the resultant proportional fair metric starts to decrease.

### III. OPTIMAL UPLINK BAND ALLOCATION

Unlike the downlink where there is a large discrepancy of transmission power between the cellular band and the TV white space, the uplink transmit power limit in the TV band for mobile devices is close to the typical cellular uplink transmission power. This difference causes completely different allocation strategies between uplink and downlink.

The original data rate for user $u \in \mathbf{U}$ on the traditional cellular uplink is

$$C(u) = \frac{W}{|\mathbf{U}|} \log\left(1 + \frac{P_c \eta_c(u)}{\frac{W}{|\mathbf{U}|} N_0}\right) = \frac{W}{|\mathbf{U}|} \log\left(1 + \frac{|\mathbf{U}| P_c \eta_c(u)}{W N_0}\right) \quad (9)$$

where $\eta_c(u)$ is the user $u$'s path loss in cellular band and $P_c$ is each user's total transmission power on the uplink, and $N_0$ is the noise spectral density.

Adopting the same analysis methodology from the downlink, we again assume that scheme 1 moves higher geometry $u_1$ to the TV band ($u_2$ with lower geometry remains in the cellular band) and scheme 2 instead moves $u_2$ to the TV band ($u_1$ remains in the cellular band). The only terms that differentiate the proportional fair metrics in (5) and (6) corresponding to the two schemes are now

$$\log\left(1 + \frac{(|\mathbf{U}_T|+1) P_T \eta_T(u_1)}{W N_0}\right) \log\left(1 + \frac{(|\mathbf{U}_c|-1) P_c \eta_c(u_2)}{W N_0}\right) \quad (10)$$

for scheme 1, and

$$\log\left(1 + \frac{(|\mathbf{U}_T|+1) P_T \eta_T(u_2)}{W N_0}\right) \log\left(1 + \frac{(|\mathbf{U}_c|-1) P_c \eta_c(u_1)}{W N_0}\right) \quad (11)$$

for scheme 2, where $P_T \approx P_c \approx 20$ dBm. It is easy to verify that (11) is greater than (10) given $d_1 < d_2$. Therefore, (6) is greater than (5), indicating that the lower geometry user should be moved to the TV band, contrary to the downlink case. It can then be concluded that the maximizer of (4), $\tilde{\mathbf{U}}_T$, on the uplink, must include the lowest geometry users and $\tilde{\mathbf{U}}_c = \mathbf{U} \setminus \tilde{\mathbf{U}}_T$ the highest geometry users.

This conclusion becomes clearer by comparing the data rate of a low geometry user with (9), should it remain in the cellular band,

$$\frac{C_c(u)}{C(u)} = \frac{\frac{W}{|\mathbf{U}_c|} \log\left(1 + \frac{|\mathbf{U}_c| P \eta_c(u)}{W N_0}\right)}{\frac{W}{|\mathbf{U}|} \log\left(1 + \frac{|\mathbf{U}| P \eta_c(u)}{W N_0}\right)} \approx \frac{\frac{W}{|\mathbf{U}_c|} \frac{|\mathbf{U}_c| P \eta_c(u)}{W N_0}}{\frac{W}{|\mathbf{U}|} \frac{|\mathbf{U}| P \eta_c(u)}{W N_0}} = 1 \quad (12)$$

where we use the fact that the SNR for low geometry users are typically low. Equation (12) indicates an important fact that low geometry users do not benefit from staying in the cellular band on the uplink even with increased bandwidth. They benefit more from the superior propagation of the TV band since these users are mainly power limited. The data rate that the low geometry user can achieve by moving over to the TV band is

$$\frac{C_T(u)}{C(u)} = \frac{\frac{W}{|\mathbf{U}_T|}\log\left(1+\frac{|\mathbf{U}_T|P\eta_T(u)}{WN_0}\right)}{\frac{W}{|\mathbf{U}|}\log\left(1+\frac{|\mathbf{U}|P\eta_c(u)}{WN_0}\right)} \quad (13)$$

$$\approx \frac{\frac{W}{|\mathbf{U}_T|}\frac{|\mathbf{U}_T|P\eta_T(u)}{WN_0}}{\frac{W}{|\mathbf{U}|}\frac{|\mathbf{U}|P\eta_c(u)}{WN_0}} = \frac{\eta_T(u)}{\eta_c(u)} \gg 1$$

which is directly proportional to the path power gain in the TV band over the cellular band. As for the high geometry users, it makes sense to leave them in the cellular band since they do not need the superior propagation of the TV band to boost up their receive power at the base station as much as the low geometry users do. The high geometry users are mainly bandwidth limited. They benefit more from the freed up bandwidth.

This conclusion leads to the important simplification of (4) to the problem of determining the size of the lowest geometry users that needs to be moved to the TV band. We can simply hypothesis-test-move the lowest geometry users to the TV band one by one from the bottom of the user list until the resultant proportional fair metric starts to decrease.

## IV. SIMULATION RESULTS

The performance of the frequency band allocation algorithm was evaluated via a cognitive cellular network simulator and the effect of utilizing the TV white space on the cellular network performance was assessed. The conventional hexagonal cellular network layout was used. Cells were sectorized with three sectors per site. Both the vertical and horizontal antenna patterns and the orientations have been considered while evaluating path losses. Users were randomly dropped into each sector. The cellular carrier frequency was 2 GHz whose propagation path loss was modeled by the Cost-231 model [4]. The TV frequency was 600 MHz and was characterized by the Hata model [4]. Full buffer traffic model with proportional-fair scheduling was assumed throughout this study. More simulation parameters are summarized in Table I.

Fig. 1 gives the downlink user throughput CDFs for both traditional and cognitive cellular networks. It is seen that all users benefit from the use of TV white space spectrum under the optimal band allocation scheme. As expected, the percentage in throughput increase of the low geometry users (cell edge users) is close to 100%. This is due to the fact that the low geometry users are maintained in the cellular band. Their spectral efficiency therefore remain the same. However, more bandwidth are freed up and available for use after the high geometry users are moved over to the TV band. This point is clearly seen from (2).

Fig. 1 also shows the uplink user throughput CDFs. It is seen that low geometry users receive a significant increase in throughput after being placed in TV band. This gain, as is predicted by (13), is fully due to the propagation gain of the TV frequency over the cellular frequency. We also see an throughput increase for high geometry users who remained in the cellular band. This gain, however, is due to the extra bandwidth evacuated by moving the low geometry users.

## V. Conclusion

In this letter, maximizing the benefit from the TV white space spectrum for the cellular network was investigated. In particular, the optimal band allocation was studied. It was shown that the optimal band allocation schemes on the downlink and uplink are quite different. On the downlink, the TV band should be allocated to highest geometry users while on the uplink lowest geometry users should be served in TV band. Based on the above analysis results, an optimal band allocation scheme for both links have been proposed. It was found that the use of the optimal band allocation not only results in an overall performance improvement but also leads to a very desirable performance gain for cell edge users both on the downlink and uplink. On the downlink, the benefit that cell edge users gain is from the extra bandwidth on the cellular band by offloading the high geometry users to the TV band. On the uplink, the edge user throughput is improved solely owing to the superior propagation property of the TV band frequency.

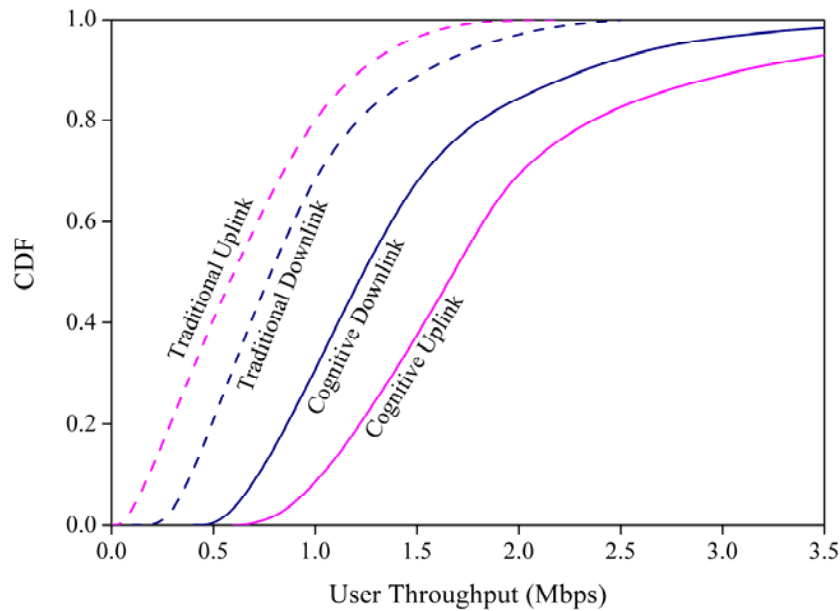

Fig. 1. User throughput CDFs for both downlink and uplink and for both the traditional and the cognitive cellular networks.

TABLE I. SIMULATION PARAMETERS

|  | Cellular | TV |
|---|---|---|
| Carrier frequency | 2 GHz | 600 MHz |
| Propagation model | Cost-231 | Hata |
| Bandwidth | 5 MHz | |
| Subcarrier number / Interval | 512 / 10 kHz | |
| Frequency plan | FFR[5][6] | |
| Base transmit power | 42 dBm | 30 dBm |
| Base antenna gain | 17 dBi | 6 dBi |
| Mobile transmit power | 20 dBm | 23 dBm |
| Mobile antenna gain | 0 dBi | -3 dBi |
| Noise figure (base / mobile) | 6 dB / 10 dB | |
| Antenna height (base / mobile) | 30 m / 2 m | |